\documentclass[letterpaper, 10 pt, conference]{ieeeconf}
\IEEEoverridecommandlockouts
\usepackage{amsmath}
\usepackage{enumerate}
\usepackage{amssymb}
\usepackage{color}
\usepackage{mathrsfs}
\usepackage{mathtools}
\usepackage{commath}
\usepackage{graphicx}
\usepackage[margin=2cm]{geometry}
\usepackage[capitalize,nameinlink,compress]{cleveref}

\newcommand{\R}{\mathbb{R}}

\newtheorem{remark}{Remark}
\newtheorem{problem}{Problem}

\crefname{appendix}{Appendix}{Appendices}
\crefname{figure}{Figure}{Figures}
\crefname{line}{line}{lines}
\crefname{claim}{Claim}{Claims}
\crefname{equation}{}{}
\crefname{problem}{Problem}{Problems}
\crefname{assumption}{Assumption}{Assumptions}

\begin{document}

\title{Advanced Safety Filter for Smooth Transient Operation of a Battery Energy Storage System}

\author{Michael Schneeberger, Florian D{\"o}rfler, Silvia Mastellone
   \thanks{Michael Schneeberger and Silvia Mastellone are with the Institute of Electrical Engineering, FHNW, Windisch, Switzerland (e-mails: michael.schneeberger@fhnw.ch, silvia.mastellone@fhnw.ch), and Florian D{\"o}rfler is with the Department of Information Technology and Electrical Engineering, ETH Z{\"u}rich, Switzerland, (e-mail: dorfler@control.ee.ethz.ch)} 
   \thanks{This work was supported by the Swiss National Science Foundation (SNSF) under NCCR Automation.}
}

\maketitle

\begin{abstract}
In this paper, we implement an advanced safety filter to smoothly limit the current of an inverter-based Battery Energy Storage System. The task involves finding suitable Control Barrier Function and Control Lyapunov Function via Sum-of-Squares optimization to certify the system's safety during grid transients. In contrast to the conventional safety filter, the advanced safety filter not only provides a safety certificate but also achieves finite-time convergence to a nominal region. Within this region, the action of the nominal control, i.e. the Enhanced Direct Power Control, remains unaltered by the safety filter. The advanced safety filter is implemented using a Quadratically Constrained Quadratic Program, providing the capability to also encode quadratic input constraints. Finally, we showcase the effectiveness of the implementation through simulations involving a load step at the Point of Common Coupling, and we compare the outcomes with those obtained using a standard vector current controller.
\end{abstract}

\section{Introduction}

A Battery Energy Storage System (BESS) enables  part of the power grid to disconnect  from the utility grid and operate independently in an islanded mode.
In this scenario, the primary objective of the BESS is to maintain grid voltage and frequency stability through the use of an inert grid-forming (GFM) control scheme.
However, when operating in grid-connected mode, the BESS's current needs to be limited during transient grid events such as faults, load steps or changes in configuration to protect the system's hardware.
In such situations, a fast response is necessary to ensure the system's safety.

In a microgrid configuration, a compelling GFM control scheme is the Enhanced Direct Power Control (EDPC) \cite{jain2020grid,rosso2021grid}.
One version of this approach, described in \cite{ferdinand2023designing}, consists of an active and reactive power controller that, together with a feed-forward term derived from a filtered voltage measurement at the Point of Common Coupling (PCC), determines the converter voltage reference.
In islanding configuration, the EDPC operates in GFM mode due to the slow dynamics of this filtered PCC voltage.
However, during  transients, faster filtered PCC voltage dynamics are temporarily permitted, even at the expense of potentially sacrificing the GFM property, to facilitate a stable return to EDPC operation.

To limit the converter current in grid-connected mode, a vector current controller can be activated based on a grid fault detection \cite{ndreko2018gfm_paper}.
However, this requires manually tuning (assuming a worst-case fault) the control gains of the current controller to respect the current limits and often causes undesired chattering behaviors resulting from the activation and deactivation of the current controller.
Current limiting strategies based on limiting the current reference, or introducing virtual impedance have been proposed \cite{schweizer2022grid,fan2022review}.
In this approach, unlike the EDPC, a vector current control is an integral part of the GFM control scheme, rendering any switching action obsolete.
None of these approaches provide a safety certification for the BESS across all states.

In this paper, we employ a \emph{safety filter} approach to limit the current and achieve a smooth transition between operation under EDPC and current limiting control.
The safety filter is obtained by solving a Quadratic Program (QP) at every state based on a Control Barrier Function (CBF) and an internal system model to ensure forward invariance of the \emph{safe set} while tracking a nominal control reference, as described in \cite{AmesCoogan:19}.
The safety filter can be integrated with a QP encoding the Control Lyapunov Function (CLF) condition, ensuring not only safety but also convergence to a specific state or set \cite{ames2016control}.
Polynomial CBF and CLF candidates, encoding the specifications of the safety filter, can be derived numerically by solving a series of Sum-of-Squares (SOS) optimization problems \cite{TanPackard:04,clark2021verification}.
Additionally, linear input constraints can be encoded as SOS constraints \cite{wang2023safety,dai2023convex}.

In our previous work, we proposed an advanced safety filter  \cite{schneeberger2024advanced}, that extends the basic version by ensuring a finite-time convergence to a forward invariant set, referred to as \emph{nominal region}, wherein the nominal controller remains undisturbed by the safety filter.
In this work we apply this concept to the BESS control scenario, where the safety filter is limiting the current while tracking the EDPC's voltage reference.
To ensure a finite-time convergence guarantee to the nominal region under stationary but unknown grid conditions, we additionally incorporate the dynamics of the filtered PCC voltage into the internal system model of the safety filter.
Utilizing the SOS-based alternating algorithm outlined in \cite{schneeberger2024advanced}, suitable CBF and CLF candidates are identified that adhere to the quadratic input constraints of the BESS.
Subsequently, the advanced safety filter is implemented using a Quadratically Constrained Quadratic Program (QCQP), providing the capability to encode the quadratic input constraints.
Finally, we demonstrate  the effectiveness of the proposed solution through simulations where  a load step is applied  at the Point of Common Coupling (PCC). The  system response is compared with the one obtained using a standard vector current controller.
This study demonstrates the applicability and efficacy of the safety filter concept in a non-trivial power electronics example beyond its traditional application in robotics.

The remainder of the paper is structured as follows:
\cref{sec:bess} describes the problem setup including the BESS, the EDPC, and the requirement on the current limiting control.
In \cref{sec:safty_filter}, we present the concept of the advanced safety filter, its properties, and the resulting SOS formulation.
Numerical results, presented in \cref{sec:numerical_results}, demonstrate the superior performance of the safety filter in comparison to a conventional vector current control \cite{ndreko2018gfm_paper}.
Finally, \cref{sec:conclusion} concludes the paper.

\section{Battery Energy Storage System} \label{sec:bess}

\subsection{Problem setup}

\begin{figure}
    \centering
    \resizebox{85mm}{!}{\includegraphics{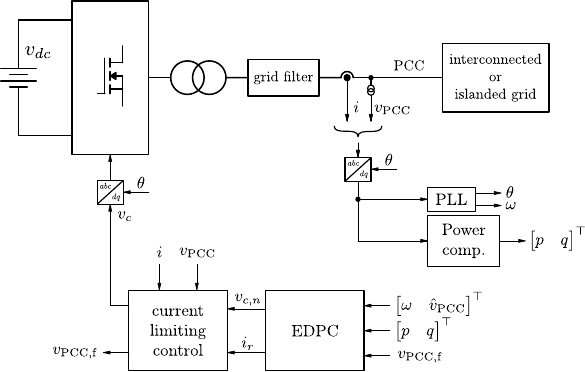}}
    \caption{
        The BESS is implemented as a three-phase power inverter system connected to the PCC via a transformer and a grid filter.
        The EDPC in series with a current limiting control ensures GFM behavior during islanded grid operation while limiting the current during grid transients.
    }
    \label{fig:overview}
\end{figure}

Consider a BESS implemented as a three-phase power inverter system connected to the PCC through a transformer, as depicted in \cref{fig:overview}.
Given that the state of charge (SOC) of the BESS's battery changes at a much slower rate than the control, we model it as a constant dc-voltage source $v_{dc}$.
A grid filter installed between transformer and PCC, along with an optimized pulse pattern (OPP), suppresses high-frequency harmonics generated by the inverter modulation.
This justifies the adoption of an averaged inverter model, where the BESS's converter voltage is expressed as the product of the dc-link voltage $v_{dc} \in \R$ and the modulation index $m \in \R^2$:
\begin{align*}
    v_c = v_{dc} m,
\end{align*}
where $m$ is constrained by $m_c^\mathrm{max} \in \R$:
\begin{align} \label{eq:max_mod_index}
    m_{c,d}^2 + m_{c,q}^2 \leq \left ( m_c^\mathrm{max} \right )^2.
\end{align}

Given that the controller's bandwidth is much smaller than the frequency band of the grid filter, we simplify the model of the grid filter by representing it as an inductance. 
Similarly, the transformer, which steps up the voltage to match the grid voltage at the PCC, is represented by an inductance. 
The current dynamics over the filter and transformer inductance $l_c$ in p.u. are expressed as
\begin{align} \label{eq:simplified_grid_model}
   \frac{l_c}{\omega_n} \begin{bmatrix} \dot i_d \\ \dot i_q \end{bmatrix} = 
   - Z_c \underbrace{\begin{bmatrix} i_d \\ i_q \end{bmatrix}}_{=:i}
   - \underbrace{\begin{bmatrix} v_{\mathrm{PCC},d} \\ v_{\mathrm{PCC},q} \end{bmatrix}}_{=:v_\mathrm{PCC}}
   + \underbrace{\begin{bmatrix} v_{c,d} \\ v_{c,q} \end{bmatrix}}_{=:v_c},
\end{align}
where $\omega_n := 2\pi 50$Hz is the nominal grid frequency, the current $i$ and the voltage $v_{\mathrm{PCC}}$ are measured, as indicated in \cref{fig:overview}, and $Z_c := J \omega l_c$ is the matrix representation of the complex filter and transformer inductance with $J = \begin{bsmallmatrix}0 & -1 \\ 1 & 0\end{bsmallmatrix}$.
The precise value of $l_c$ is determined during the commissioning of the BESS.

Operating the BESS in both grid-connected and islanding modes, requires suitable control strategies that meet grid stability and load requirements, while respecting the constraints and preserve the safety of the battery and converter components.
The BESS control architecture consist of two main components.
The slow outer loop, in form of the EDPC, is responsible for meeting the grid stability and load requirements, while the fast inner loop, in form of a current limiting controller, is designed to limit the current during transients.
Although both control loops are designed in continuous-time, they are implemented in discrete-time with sufficiently fast task cycles to preserve the controllers performance.

\subsection{Enhanced Direct Power Control} \label{sec:direct_power_control}

\begin{figure}
    \centering
    \resizebox{85mm}{!}{\includegraphics{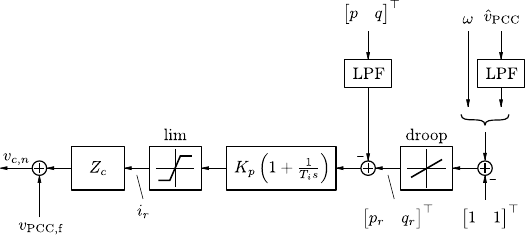}}
    \caption{
        The Enhanced Direct Power Control (EDPC) directly generates a converter voltage reference $v_{c,n}$ from a power control loop, integrating frequency and amplitude droop characteristics.
    }
    \label{fig:direct_power_control}
\end{figure}

The EDPC, taken from \cite{jain2020grid,rosso2021grid,ferdinand2023designing} and illustrated in \cref{fig:direct_power_control}, consists of an active and reactive power loop, achieving a slow time constant through the utilization of a low-pass filter (LPF) in the measurement of active power $p$ and reactive power $q$.
The active and reactive power references are determined by a frequency droop $D_f \in \R$ and a voltage droop $D_v \in \R$:
\begin{align*}
    p_r &= D_f ( \omega - 1 ) \\
    q_r &= D_v ( \hat v_{\mathrm{PCC},lp} - 1),
\end{align*}
where $\omega$ is the angular frequency of the PLL in p.u., and $\hat v_{\mathrm{PCC},lp}$ is the low-pass filtered PCC voltage amplitude.
The active and reactive power control, depicted in \cref{fig:direct_power_control}, are based on the PI controller
\begin{align*}
    K_p  \left ( 1 + \frac{1}{T_is} \right ),
\end{align*}
characterized by a proportional gain $K_p \in \R$ and integral time constant $T_i \in \R$.
This controller transforms the active and reactive power error into a (nominally unlimited) current reference.
A current reference limiter $\| i_r \|^2 = i_{r,d}^2 + i_{r,q}^2 \leq (i_r^\mathrm{lim})^2$ for some current limit $i_r^\mathrm{lim} < i^\mathrm{max}$ ensures a stable steady-state current below the maximum allowed current \cref{eq:allowable_current}.
A cross-coupling term $Z_c$ computes the voltage drop over the transformer and grid filter impedance.
Finally, a filtered PCC voltage $v_{\mathrm{PCC},f}$ is added to the voltage drop, producing the BESS's voltage reference
\begin{align} \label{eq:nomincal_vc_controller}
    v_{c,n}(i_r, v_{\mathrm{PCC},f}) := Z_c i_r + v_{\mathrm{PCC},f}.
\end{align}
The dynamical system characterizing the filtered PCC voltage is given as
\begin{align} \label{eq:filtered_pcc_voltage_dynamic}
    \underbrace{\begin{bmatrix} \dot v_{\mathrm{PCC},f,d} \\ \dot v_{\mathrm{PCC},f,q} \end{bmatrix}}_{=: \dot v_{\mathrm{PCC},f}} = \underbrace{\begin{bmatrix} \alpha_d \\ \alpha_q \end{bmatrix}}_{=:\alpha},
\end{align}
where $\alpha \in \R^2$ is the input of this system.
To achieve GFM behavior, the filtered PCC voltage must exhibit characteristics of a low-pass filter with time constant $\tau \approx 100ms$, as described by
\begin{align} \label{eq:filtered_pcc_voltage_nominal_change}
     \alpha_n(v_\mathrm{PCC}, v_{\mathrm{PCC},f}) := \frac{v_\mathrm{PCC} - v_{\mathrm{PCC},f}}{\tau}.
\end{align}
Only during transients, it is allowed to change the behavior of the filtered PCC voltage.

\subsection{The problem of current limiting control} \label{sec:current_limiting_control}

The current limiting control at the output of the EDPC, as depicted in \cref{fig:overview}, ensures that the current remains within the maximum allowed boundaries, characterized by $i^\mathrm{max} \in \R$:
\begin{align} \label{eq:allowable_current}
    i_{d}^2 + i_{q}^2 \leq (i^\mathrm{max})^2.
\end{align}
This is achieved by adjusting the EDPC's voltage reference $v_{c,n}$ \cref{eq:nomincal_vc_controller} to generate a safe voltage reference $v_c$.
Furthermore, the dynamics governing the filtered PCC voltage $v_{\mathrm{PCC},f}$ in \cref{eq:filtered_pcc_voltage_dynamic,eq:filtered_pcc_voltage_nominal_change} can be temporarily adjusted during a current transient, facilitating a stable return to EDPC operation.

\begin{remark}
    \textit{
        During islanding mode, the BESS is assumed to be sufficiently dimensioned to withstand transients, without having to resort to a fast current controller.
        This assumption is crucial, as the stability of an islanded grid consisting purely of grid-following (GFL) converters cannot be assured otherwise.
        Hence, the current limiting controller is solely necessary when operating within an interconnected grid, permitting the BESS to briefly adopt a fast GFL mode without compromising grid stability.
    }
\end{remark}

The design of the current limiting control and later of the advanced safety filter requires knowledge of the grid model, which in practice can be of arbitrarily complexity and subject to change over time.
Moreover, the original models used for the  EDPC have an unrequired degree of complexity.
Therefore, we propose the following  simplifying assumptions:
\begin{itemize}
    \item  Given that the discrete-time implementation of the safety filter utilizes updated $v_\mathrm{PCC}$ measurements within each task cycle, we adopt a simplified reduced-order grid model based on the current dynamics \cref{eq:simplified_grid_model}, where $v_\mathrm{PCC}$ is assumed to be constant in dq reference frame.
    \item Given that the EDPC operates with a  slower time constant of approximately 100ms compared to the current limiting control's time constant of around 1ms, we propose a simplified reduced-order EDPC model based on \cref{eq:nomincal_vc_controller}, where the current reference $i_r$ is assumed to be constant.
\end{itemize}
Under those assumptions, we can formulate the problem statement as follows:
\begin{problem} \label{prob:statement}
    Given the dynamical system described in  \cref{eq:simplified_grid_model,eq:filtered_pcc_voltage_dynamic}, with a constant voltage source $v_\mathrm{PCC}$, a set of allowable states \cref{eq:allowable_current}, an input constraint set \cref{eq:input_constraint_set}, and a nominal controller described in  \cref{eq:nomincal_vc_controller,eq:filtered_pcc_voltage_nominal_change}, with a constant current reference $i_r$, the task is to  design an advanced safety filter, as described in \cite{schneeberger2024advanced}, that ensures stable and safe behavior of the BESS during grid events.
\end{problem}

A standard method to solve \cref{prob:statement} is to employ a vector current control \cite{ndreko2018gfm_paper}.
This involves a PI controller integrated with an anti-windup mechanism to track a safe current reference within the dq reference frame.
However, this leads to chattering effects, abrupt behavior, and possibly system shutdown in case of protection function activation.

\section{Advanced safety filter} \label{sec:safty_filter}

\begin{figure}
    \centering
    \resizebox{75mm}{!}{\includegraphics{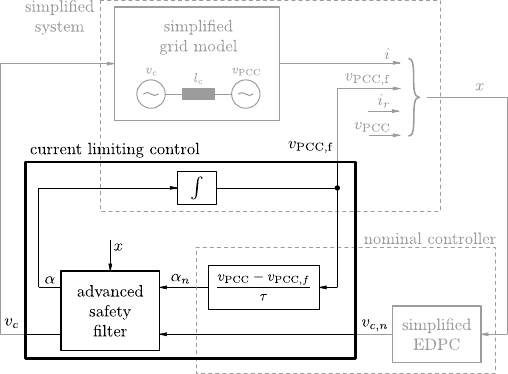}}
    \caption{
        The advanced safety filter ensures safe operation with respect to the maximum allowed current limits \cref{eq:allowable_current}.
        Additionally, it guarantees finite-time convergence to the nominal region, where the nominal control input $u_n^\top = \begin{bmatrix}
            v_{c,n}^\top & \alpha_n^\top
        \end{bmatrix}$ is -- assuming simplified system and EDPC control dynamics -- not perturbed by the advanced safety filter.
        The dynamics of the filtered PCC voltage $v_{\mathrm{PCC},f}$ is also taken into account to ensure the existence of the nominal region.
    }
    \label{fig:safety_filter_setup}
\end{figure}

To overcome the limitations of the state of the art vector current controller, we propose an alternative solution to the problem of overcurrent phenomena happening during transients.
Our approach allows to promptly respond to grid events while guaranteeing a smooth transitioning between EDPC and current limiting control operation.
This entails designing an advanced safety filter setup, as shown in \cref{fig:safety_filter_setup}, to enforce the state and input constraints while ensuring finite-time convergence to the EDPC operation.

Consider the control system composed of the simplified grid model derived from \cref{eq:simplified_grid_model}, the filtered PCC voltage dynamics derived from \cref{eq:filtered_pcc_voltage_dynamic}, and the stationary dynamics of the current reference $i_r$ and PCC voltage $v_\mathrm{PCC}$ as follows:
\begin{align} \label{eq:bess_control_system}
    \dot x = \underbrace{\begin{bmatrix}
        -(\omega_n / l_c) \left( Z_c i + v_\mathrm{PCC} \right) \\
        \mathrm{0} \\ \mathrm{0} \\ \mathrm{0}
    \end{bmatrix}}_{=:f(x)} + \underbrace{\begin{bmatrix}
        (\omega_n / l_c) \mathrm{I}_2 &  \mathrm{0} \\  \mathrm{0} & \mathrm{I}_2 \\ \mathrm{0} & \mathrm{0} \\ \mathrm{0} & \mathrm{0}
    \end{bmatrix}}_{=:G(x)} \underbrace{\begin{bmatrix}
        v_c \\ \alpha
    \end{bmatrix}}_{=:u},
\end{align}
where 
$x^\top := \begin{bmatrix}i^\top & v_{\mathrm{PCC},f}^\top & i_r^\top & v_\mathrm{PCC}^\top\end{bmatrix}$ is the state of the system.
The reduced-order model of the inverter current $i$ and the filtered PCC voltage $v_{\mathrm{PCC},f}$ are derived from the specifications outlined in \cref{sec:current_limiting_control}.
The current reference $i_r$ and PCC voltage $v_\mathrm{PCC}$, although considered constant, are included in the system dynamics to render the safety filter parametric in these variables.
The set of allowed states is defined by the maximum allowed current: 
\begin{align} \label{eq:allowable_set}
    \mathcal X_a = \left \{ x \mid 
    \text{subject to \cref{eq:allowable_current}} \right \}.
\end{align}
The input constraint set resulting from \cref{eq:max_mod_index} is given as:
\begin{align} \label{eq:input_constraint_set}
    \mathcal U = \left \{ u \mid v_{c,d}^2 + v_{c,q}^2 \leq \left ( m_c^\mathrm{max} \right )^2 \right \},
\end{align}
where dc-link voltage is assumed to be $v_{dc} = 1$ p.u.
All parameter values can be found in \cref{tab:param_values}.

\subsection{Control Barrier and Lyapunov-like Function} \label{sec:cbf_clf}

\begin{figure}
    \centering
    \resizebox{70mm}{!}{\includegraphics{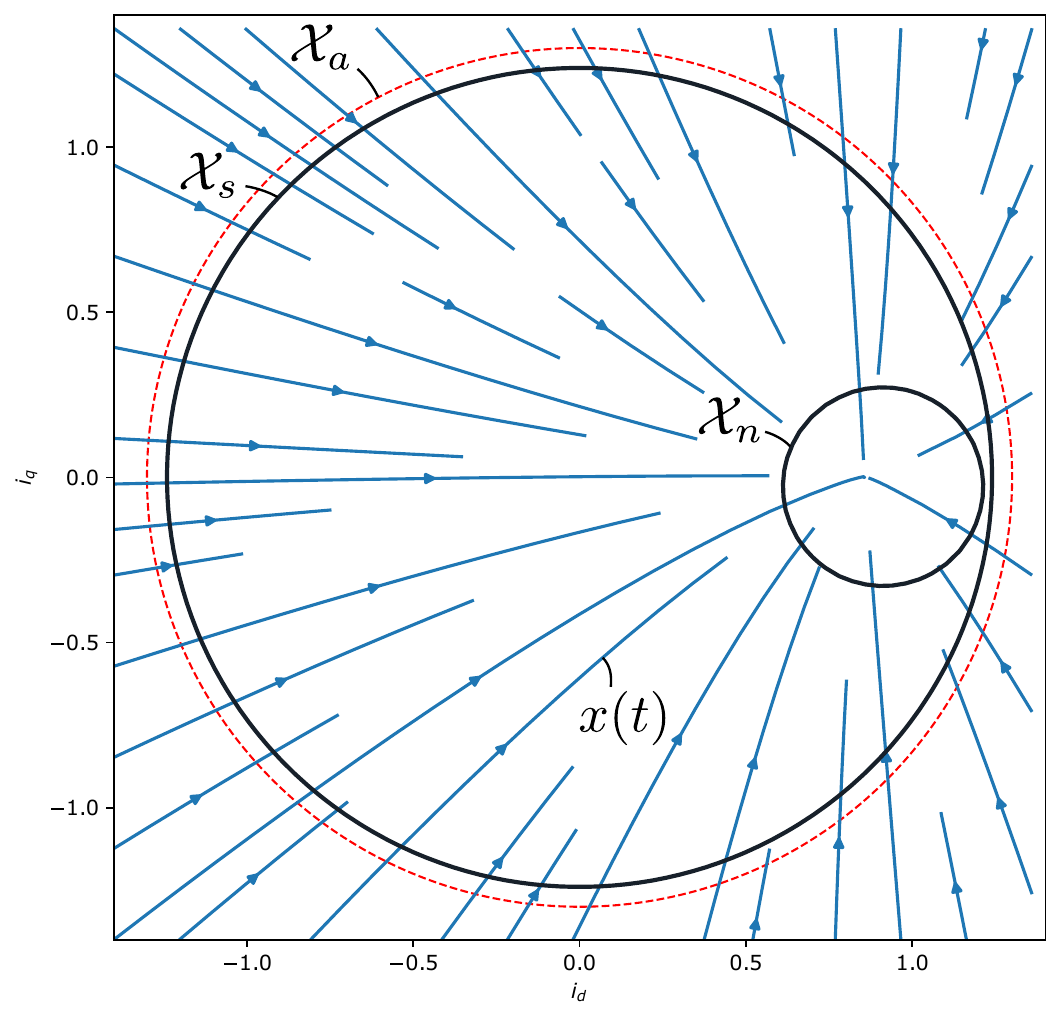}}
    \caption{
        The polynomial CBF $B(x)$ and CLF $V(x)$, along with their corresponding safe set $\mathcal X_s$, which must be contained within the allowed set of states $\mathcal X_a$, and nominal region $\mathcal X_n$, are computed using SOS optimization.
        The vector field $\dot x = f(x) + G(x) u_{sos}(x)$, projected to $(i_{d}, i_q)$ coordinates setting $(i_{r,d}, i_{r,q}) = (0, 1)$, is depicted in blue.
        Here, the polynomial $u_{sos}(x)$ is utilized to ensure compatibility between the CBF and CLF.
    }
    \label{fig:safe_and_nominal_set}
\end{figure}

The construction of the advanced safety filter relies on the development of a CBF and CLF, which are reviewed in this subsection.
We consider an abstract polynomial control system, affine in the control action $u \in \R^m$, and given as
\begin{align} \label{eq:control_system}
   \dot x = f(x) + G(x) u,
\end{align}
where $x \in \R^n$ is the state, $f(x) \in \left( R[x] \right)^n$ is a polynomial vector, and $G(x) \in \left( R[x] \right)^{n \times m}$ is a polynomial matrix.
System~\cref{eq:control_system} with a polynomial state feedback control policy $u_s(x) \in \left( R[x] \right)^m$ results in the closed-loop system
\begin{align} \label{eq:closed_loop_system}
   \dot x = f(x) + G(x) u_s(x).
\end{align}
A subset $\mathcal X_s \subseteq \R^n$ is called \emph{forward invariant} (cf. \cite[Theorem 4.4]{Khalil:02}) with respect to system \cref{eq:closed_loop_system} if for every $x(0) \in \mathcal X$, $x(t) \in \mathcal X$ for all $t \geq 0$.
A system \cref{eq:closed_loop_system} is called \emph{safe} (cf. \cite{WielandAllgower:07}) w.r.t. an allowable set $\mathcal X_a \subseteq \R^n$ and the safe set $\mathcal X_s \subseteq \R^n$, if $\mathcal X_s$ is forward invariant and $\mathcal X_s \subseteq \mathcal X_a$.

Safety of a control system \cref{eq:control_system} can be asserted with the existence of a differentiable function $B: \R^n \to \R$ such that for all states 
$x \in \partial B(x)$ there exists $u \in \mathcal U$:
\begin{align} \label{eq:cbf_condition}
    \nabla B(x)^\top \left ( f(x) + G(x) u \right ) \leq 0.
\end{align}
Such a function $B(x)$ is called a Control Barrier Function (CBF), and its zero-sublevel set $\mathcal X_s := \{ x \mid B(x) \leq 0 \}$, to be contained in $\mathcal X_a$ \cref{eq:allowable_set}, defines the safe set.

The advanced safety filter $u_s(x)$ ensures a finite-time convergence to the nominal region using
a differentiable function $V: \R^n \to \R$ with a strictly positive dissipation rate $d(x) > 0$ such that for all $x \in \{ x \mid V(x) \leq x \leq B(x) \}$ there exists $u \in \mathcal U$ (cf. \cite{schneeberger2024advanced}):
\begin{align} \label{eq:clf_condition}
    \nabla V(x)^\top \left ( f(x) + G(x) u \right ) + d(x) \leq 0.
\end{align}
Such a function $V(x)$ is referred to as a Control Lyapunov-Like Function (CLF), and its zero-sublevel set $\mathcal X_n := \{ x \mid V(x) \leq 0 \} \subsetneq \mathcal X_s$ defines the nominal region.
To ensure compatibility with the nominal controller $u_n(x)$, we furthermore require that \cref{eq:clf_condition} holds on the boundary of $\mathcal X_n$ with $u = u_n(x)$, see \cite{schneeberger2024advanced} for details, that is for all $x \in \partial X_n$:
\begin{align} \label{eq:clf_condition_un}
    \nabla V(x)^\top \left ( f(x) + G(x) u_n(x) \right ) + d(x) \leq 0.
\end{align}
With these conditions established, we can formulate the QCQP defining the advanced safety filter.

\subsection{Quadratically Constrained Quadratic Program}

When provided with a CBF $B(x)$ and a CLF $V(x)$ that satisfy conditions \cref{eq:cbf_condition,eq:clf_condition,eq:clf_condition_un}, the advanced safety filter can be implemented using the QCQP
\begin{align} \label{eq:qp_based_controller}
   \begin{array}{llll}
      u_s(x) := &\underset{u \in \mathcal U}{\mbox{min}} & \| u_n(x) - u \|^2 \\
      &\mbox{s.t.} & C(x) u + b(x) \leq 0,
   \end{array}
\end{align}
encoding the input constraint set $\mathcal U$ in \cref{eq:input_constraint_set} as quadratic constraints of the QCQP.
The nominal controller
\begin{align} \label{eq:nominal_controller}
    u_n(x) = \begin{bmatrix}
        Z_c i_r + v_{\mathrm{PCC},f} \\
        \tau (v_{\mathrm{PCC}} - v_{\mathrm{PCC},f})
    \end{bmatrix}
\end{align}
incorporates the behavior of the simplified EDPC \cref{eq:nomincal_vc_controller} and the nominal rate of the filtered PCC voltage \cref{eq:filtered_pcc_voltage_nominal_change}.
The state-dependent matrix $C(x) \in \R^{2 \times 4}$ and vector $b(x) \in \R^2$ encode the CBF and CLF conditions \cref{eq:cbf_condition,eq:clf_condition} as follows:
\begin{subequations} \label{eq:def_c_and_b}
   \begin{align} \label{eq:def_c}
      C(x) &:= \begin{bmatrix} 
        \nabla B(x)^\top G(x) \\
        \nabla V(x)^\top G(x)
      \end{bmatrix} \\
      \label{eq:def_b}
      b(x) &:= \begin{bmatrix}
        \nabla B(x)^\top f(x) \\
        \nabla V(x)^\top f(x) + d(x)
      \end{bmatrix} - \begin{bmatrix}
         r_0(x) \\ r_1(x)
      \end{bmatrix},
   \end{align}
\end{subequations}
where $r_0(x)$ and $r_1(x)$ are state-dependent slack variables that are non-positive for all states $x$, where the CBF or CLF conditions are applicable.
By choosing them smoothly, we can achieve a Lipschitz-continuous QCQP-based controller, see \cite[Theorem 2]{schneeberger2024advanced} for details.

\subsection{Sum-of-Squares Optimization} \label{sec:sum_of_squares}

A polynomial CLF and CBF can be found numerically using Sum-of-Squares (SOS) optimization.
The CBF and CLF conditions \cref{eq:cbf_condition,eq:clf_condition,eq:clf_condition_un}, along with the input constraint set \cref{eq:input_constraint_set} and containment conditions $\mathcal X_n \subsetneq \mathcal X_s \subsetneq \mathcal X_a$, are represented as SOS constraints.
This is achieved by substituting a polynomial controller $u_\mathrm{SOS}(x)$ for the input vector $u$ in \cref{eq:cbf_condition,eq:clf_condition}.
The SOS constraints encoding \cref{eq:cbf_condition,eq:clf_condition,eq:clf_condition_un} are obtained by applying Putinar's Positivstellensatz \cite[Eq. (27)]{schneeberger2024advanced}:
\begin{subequations} \label{eq:sos_constr_ineq}
   \begin{align}
      \begin{split} \label{eq:sos_constr_ineq_cbf}
         \nabla B^\top \left ( f + G u_\mathrm{SOS} \right ) + \gamma_B B + \gamma_1 f_\mathrm{op} &\in - \Sigma[x] 
      \end{split} \\
      \begin{split} \label{eq:sos_constr_ineq_clf}
         \nabla V^\top \left ( f + G u_\mathrm{SOS} \right ) + d + \gamma_V V - \gamma_r B & \\
         + \gamma_2 f_\mathrm{op} &\in - \Sigma[x]
      \end{split} \\
      \begin{split} \label{eq:sos_constr_ineq_nom}
         \nabla V^\top \left ( f + G u_{n} \right ) + d + \gamma_n V + \gamma_3 f_\mathrm{op} &\in - \Sigma[x],
      \end{split}
   \end{align}
\end{subequations}
for some $\gamma_B, \gamma_n \in R[x]$, and $\gamma_V, \gamma_r, \gamma_1, \gamma_2, \gamma_3 \in \Sigma[x]$, where $R[x]$, resp. $\Sigma[x]$, refers to the set of scalar polynomials, resp. SOS polynomials, in variables $x \in \R^n$.
An operational region $\mathcal X_\mathrm{op}$, defined by the scalar function $f_\mathrm{op} \in R[x]$, is established to encode the considered operational ranges, where conditions \cref{eq:cbf_condition,eq:clf_condition,eq:clf_condition_un} are expected to hold.

To improve the control robustness with respect to time discretization effect, uncertainty, and the use of the simplified model \cref{eq:simplified_grid_model}, we propose the following measures:
\begin{itemize}
    \item Introduce a margin between $\mathcal X_a$ and $\mathcal X_s$.
    \item Adjust the nominal controller $u_n(x)$ in \cref{eq:sos_constr_ineq_nom} to expand the volume of the nominal region $\mathcal X_n$.
    A larger nominal region improves stability by taking advantage of the stable behavior of nominal control near the equilibrium.
\end{itemize}
The resulting robust advanced safety filter can maintain safety guarantees even under more complex grid models and demonstrates significantly smoother behavior as it transitions into the nominal region and eventually steady-state operations.

\section{Numerical Results} \label{sec:numerical_results}

\subsection{Finding CBF and CLF via SOS optimization}

The safe set $\mathcal X_s$ and nominal region $\mathcal X_n$, as shown in \cref{fig:safe_and_nominal_set}, are computed by using polynomials $B(x)$ and $V(x)$ of degree two.
Considering the linearity of the system dynamics \cref{eq:bess_control_system} and the quadratic nature of the constraints, it is possibly expected that low-degree polynomial candidates are sufficient to find solutions to the SOS problem involving \cref{eq:sos_constr_ineq}, though it's not perfectly obvious since \cref{eq:sos_constr_ineq} are still coupled quadratic equations.
The CBF is selected such that $B(0) = -1$ to ensure numerical stability.
The dissipation function in \cref{eq:clf_condition} is chosen to achieve the desired convergence to the nominal region.
The operational region
\begin{align*}
    \mathcal X_\mathrm{op} = \{ x \mid & (i_{r,d}/i_r^\mathrm{lim})^2 + (i_{r,q}/i_r^\mathrm{lim})^2 \leq 1 \\
    & (v_{\mathrm{PCC},d}/0.1)^2 + (v_{\mathrm{PCC},q}/0.1)^2 \leq 1 \}
\end{align*}
is selected to include the considered $i_r$ and $v_\mathrm{PCC}$ range.
To enhance robustness with a current margin given by $i^\mathrm{lim} < i^\mathrm{max}$, as described in \cref{sec:sum_of_squares}, and to maintain compactness of the safe set w.r.t. $i$ and $v_{\mathrm{PCC},f}$, we enforce the following subset condition:
\begin{align*}
    \mathcal X_s \subseteq \{ x \mid & (i_{d}/i^\mathrm{lim})^2 + (i_{q}/i^\mathrm{lim})^2 \leq 1 \\
    & (v_{\mathrm{PCC},f,d}/20)^2 + (v_{\mathrm{PCC},f,q}/20)^2 \leq 1 \} \subsetneq \mathcal X_a.
\end{align*}
Furthermore, the nominal controller \cref{eq:nominal_controller} used in the SOS optimization \cref{eq:sos_constr_ineq} is made more aggressive as
\begin{align*}
    u_{n}(x) = \begin{bmatrix}
        0.2 (i_r - i) + Z_c i_r + v_{\mathrm{PCC},f} \\
        10 \cdot \tau (v_{\mathrm{PCC}} - v_{\mathrm{PCC},f})
    \end{bmatrix}.
\end{align*}
A more aggressive nominal controller, while potentially sacrificing the GFM property, can enhance the finite-time convergence and forward invariance guarantees encoded in \cref{eq:sos_constr_ineq_clf} and \cref{eq:sos_constr_ineq_nom}, resulting in a larger nominal set $\mathcal X_n$.

\subsection{Load step simulation}

\begin{figure}
    \centering
    \resizebox{85mm}{!}{\includegraphics{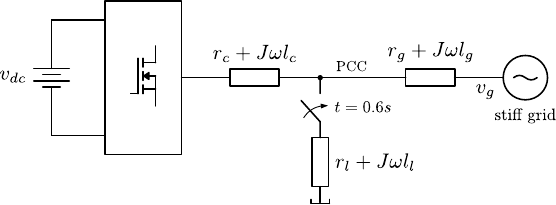}}
    \caption{
        Simulations are conducted by switching a load to the PCC, necessitating the limitation of the BESS's current.
    }
    \label{fig:load_step}
\end{figure}

\begin{table}[!t]
   \caption{Simulation parameters}
   \label{tab:param_values}
   \centering
   \begin{tabular}{|c|c|c|c|} 
      \hline
      Parameter & Symbol & Value & Unit \\
      \hline \hline
      Transformer impedance & $l_c$, $r_c$ & 0.16, 0.01 & p.u. \\
      Line impedance & $l_g$, $r_g$ & 0.016, 0.001 & p.u. \\
      Load impedance & $l_l$, $r_l$ & 0.016, 0.001 & p.u. \\
      Grid frequency & $f_g$ & 1.02 & p.u. \\
      Current reference limit & $i_r^\mathrm{lim}$ & 1.18 & p.u. \\
      Current limit & $i^\mathrm{lim}$ & 1.24 & p.u. \\
      Maximum current & $i^\mathrm{max}$ & 1.30 & p.u. \\
      Dc-link voltage & $v_{dc}$ & 1 & p.u. \\
      Modulation limit & $m_c^\mathrm{max}$ & 1.2 & p.u. \\
      Frequency droop & $D_f$ & -0.02 & p.u. \\
      Voltage droop & $D_v$ & 0.05 & p.u. \\
      Proportional gain & $K_p$ & 0.45 & p.u. \\
      Integral time constant & $T_i$ & 40 & ms \\
      Filter time constant & $\tau$ & 1 & ms \\
      \hline
   \end{tabular}
\end{table}

In this section, we investigate the behavior of the EDPC in combination with a current limiting control during a load step applied at the PCC at $t=0.6$s, as illustrated in \cref{fig:load_step}.
The QCQP-based controller $u_s(x)$ in \cref{eq:qp_based_controller}, theoretically continuous in time, is implemented discretely with a sampling time of $T_s = 200 \mu s$.
The slack variables in \cref{eq:def_c_and_b} are chosen as follows:
\begin{align*}
    r_0(x) &:= - \gamma_B(x) B(x) \quad
    r_1(x) := - \gamma_V(x) V(x), 
\end{align*}
where $\gamma_B$ and $\gamma_V$ are taken from the SOS problem \cref{eq:sos_constr_ineq}.
The performance of the safety filter implementation is evaluated against the standard vector current control \cite{ndreko2018gfm_paper}.
In this method, the vector current control is activated whenever the current amplitude $\| i \| = \sqrt{i_d^2 + i_q^2} \geq i^\mathrm{lim}$, with a deactivation hysteresis of $0.03$ p.u.

\begin{figure}
    \centering
    \resizebox{85mm}{!}{\includegraphics{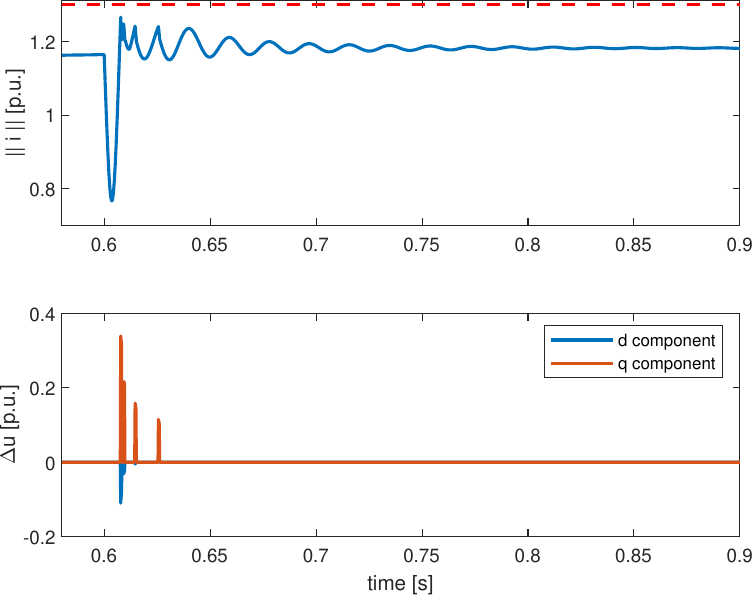}}
    \caption{
        The vector current control \cite{ndreko2018gfm_paper} is activated when the current amplitude $\| i \| \geq i^\mathrm{lim}$, with a deactivation hysteresis of $0.03$ p.u.
        This results in abrupt interventions and a chattering behavior until the current oscillations are sufficiently reduced to remain below the current amplitude limit.
    }
    \label{fig:sim_cc_f1d02}
\end{figure}

\begin{figure}
    \centering
    \resizebox{85mm}{!}{\includegraphics{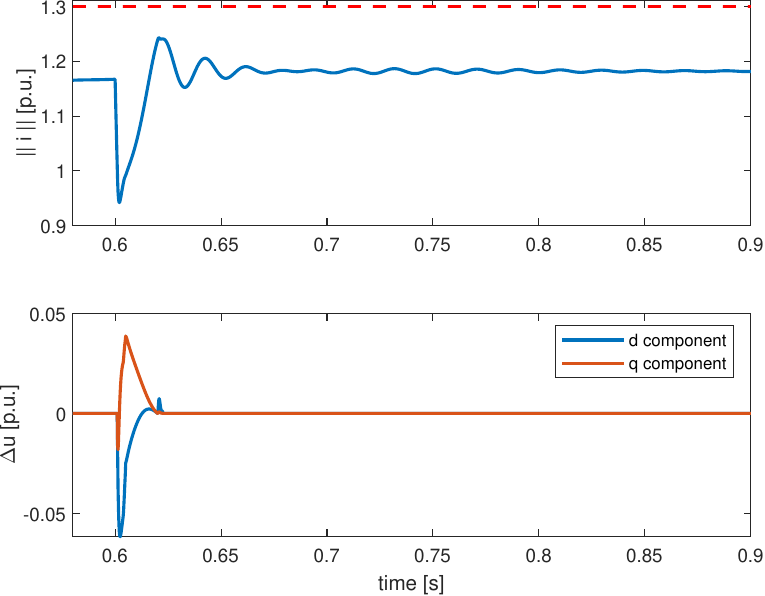}}
    \caption{
        The designed safety filter not only smoothly limits the current amplitude $\| i \|$ but also mitigates the current dip occurring immediately after the load step at $t=0.6$s.
    }
    \label{fig:sim_sf_f1d02}
\end{figure}

The simulations in \cref{fig:sim_cc_f1d02,fig:sim_sf_f1d02} show a load step using a vector current control and a safety filter with the frequency of $v_g$ set to $f_g = 1.02$ p.u.
The intervention of the safety filter and the vector current control $\Delta u$ is defined as:
\begin{align*}
    \Delta u(x) = u_s(x) - u_n(x).
\end{align*}
Unlike vector current control, which tends to exhibit abrupt interventions and a chattering behavior even with the use of a hysteresis, the safety filter demonstrates a significantly smoother operation.
This smooth operation is achieved by means of two key conditions.
Firstly, the finite-time convergence condition imposed by CLF dynamically adjusts the nominal control action to mitigate the current dip immediately after the load step.
Secondly, the safety condition imposed by CBF dynamically adjusts the nominal control action as the current approaches the allowed current amplitude $i^\mathrm{max}$ \cref{eq:allowable_current}.
The intervention of the safety filter, as expressed by $\Delta u$, exhibits a significant smaller magnitude and a smoother nature when compared to the vector current control.

\section{Conclusions} \label{sec:conclusion}

In this paper, we demonstrated our advanced safety filter and extended the application of safety filters to a non-trivial power electronics example beyond their traditional domain in robotics.
The finite-time convergence guarantee of the advanced safety filter played a key role in finding the balance between ensuring the system's safety and maintaining the behavior of the nominal controller.
Specifically, we successfully implemented an advanced safety filter capable of effectively limiting the current during a load step while preserving its GFM behavior, embodied by the EDPC, during stationary grid conditions.

\bibliographystyle{unsrt}
\bibliography{main}

\end{document}